\documentclass[fleqn]{mn2e}
\usepackage{times}
\usepackage{amssymb}
\usepackage{amsmath}
\usepackage{graphicx}

\newcommand{\bfe}{{\boldsymbol{e}}}
\newcommand{\bfk}{{\boldsymbol{k}}}
\newcommand{\bfr}{{\boldsymbol{r}}}
\newcommand{\bfv}{{\boldsymbol{v}}}
\newcommand{\bfx}{{\boldsymbol{x}}}
\newcommand{\txd}{{\text{d}}}
\newcommand{\txp}{{\text{p}}}
\newcommand{\txV}{{\text{V}}}
\renewcommand{\leq}{\leqslant}

\DeclareMathOperator{\erf}{erf}
\DeclareMathOperator{\sgn}{sgn}

\title[The observed kinematics of dusty disc galaxies]%
{Radiative transfer
in disc galaxies -- III. The observed kinematics of dusty disc galaxies}

\author[Baes et al.]{Maarten Baes$^{1,2,}$\thanks{Postdoctoral Fellow
of the Fund for Scientific Research, Flanders, Belgium
(F.W.O.-Vlaanderen)}, Jonathan I. Davies$^2$, Herwig Dejonghe$^1$,
Sabina Sabatini$^2$, \newauthor Sarah Roberts$^2$, Rhodri Evans$^2$,
Suzanne M. Linder$^2$, Rodney Smith$^2$ and \newauthor W.\,J.\,G. de Blok$^2$ \\ $^1$Sterrenkundig
Observatorium, Universiteit Gent, Krijgslaan 281-S9, B-9000 Gent,
Belgium \\ $^2$Department of Physics and Astronomy, Cardiff
University, 5 The Parade, Cardiff CF24\,3YB, Wales, UK}

\begin{document}

\maketitle

\begin{abstract}
We present SKIRT (Stellar Kinematics Including Radiative Transfer), a
new Monte Carlo radiative transfer code that allows the calculation of
the observed stellar kinematics of a dusty galaxy. The code
incorporates the effects of both absorption and scattering by
interstellar dust grains, and calculates the Doppler shift of the
emerging radiation exactly by taking into account the velocities of
the emitting stars and the individual scattering dust grains. The code
supports arbitrary distributions of dust through a cellular approach,
whereby the integration through the dust is optimized by means of a
novel efficient trilinear interpolation technique.

We apply our modelling technique to calculate the observed kinematics
of realistic models for dusty disc galaxies. We find that the effects
of dust on the mean projected velocity and projected velocity
dispersion are severe for edge-on galaxies. For galaxies which deviate
more than a few degrees from exactly edge-on, the effects are already
strongly reduced. As a consequence, dust attenuation cannot serve as a
possible way to reconcile the discrepancy between the observed shallow
slopes of the inner rotation curves of LSB galaxies and the
predictions of CDM cosmological models. For face-on galaxies, the
velocity dispersion increases with increasing dust mass due to
scattering, but the effects are limited, even for extended dust
distributions. Finally, we show that serious errors can be made when
the individual velocities of the dust grains are neglected in the
calculations.
\end{abstract}

\begin{keywords}
dust, extinction -- galaxies: kinematics and dynamics -- galaxies:
spiral -- radiative transfer
\end{keywords}

\section{Introduction}

During the last decade of the past century there has been a vivid
discussion about the opacity of spiral galaxies. This discussion was
initiated by Disney, Davies \& Phillipps~(1989) and Valentijn~(1990),
who countered the conventional view that spiral galaxies are optically
thin over their entire optical discs (Holmberg~1958; de Vaucouleurs,
de Vaucouleurs \& Corwin~1976; Sandage \& Tammann~1981). This issue
has not yet been settled, although an overall consensus seems to be
emerging that spiral galaxies are, in general, at least moderately
optically thick within their optical disc, i.e.\ that they have
central face-on optical depth of at least unity. We will not repeat a
detailed overview of the different points of view and the large
variety of observational tests, and refer to Davies \&
Burstein~(1995), Xilouris et al.~(1997), Kuchinski et al.~(1998) and
Calzetti~(2001) for excellent reviews. We just wish to point out an
important observational constraint which, in our opinion, is not
always given enough attention: the far-infrared (FIR) emission of
spiral galaxies. This emission originates from the absorption of
starlight by dust grains, which is thermally re-emitted at longer
wavelengths. The most recent results show that about 30 per cent of
the bolometric luminosity of late-type galaxies is FIR emission by
interstellar dust (Popescu \& Tuffs 2002). Notice that this figure
refers to normal quiescent spiral galaxies; special classes of
galaxies such as starburst galaxies emit even higher fractions of
their energy at FIR wavelengths. These facts demonstrate quite
convincingly that interstellar dust forms an important constituent of
spiral galaxies.

The physical processes of absorption and scattering by dust grains
should hence be taken into account when interpreting and modelling the
observed optical properties of galaxies. We would like to stress that
{\em all} optical observables will to some extent be affected by dust,
including the observed kinematics. Indeed, when interstellar dust
grains absorb or scatter photons, the kinematical information
contained within these photons, in the form of the Doppler shift, will
also be absorbed or scattered. It is important to realize that most of
the kinematical data of galaxies are optical measurements. Firstly,
stellar kinematics are always measured from optical absorption
lines. In principle, the attenuation effects of interstellar dust
could be minimized by using the sensitive $^{12}$CO absorption feature
at 2.29~$\mu$m (Gaffney, Lester \& Doppmann 1995). Measuring stellar
kinematics at near-infrared (NIR) wavelengths is, however,
observationally extremely difficult due to the high sky surface
brightness. Secondly, the vast majority of spatially resolved rotation
curves of spiral galaxies have been measured using optical emission
lines (usually the H$\alpha$ line). They can also be measured using
the H{\sc i} or CO lines, where dust extinction is not an
issue. However, these methods are observationally much more expensive
than the more straightforward H$\alpha$ observations.

Currently, the only way to investigate the effects of dust attenuation
on the observed kinematics of galaxies is by means of detailed
radiative transfer modelling. Many different approaches exist to
handle the radiative transfer problem, and various teams have adopted
different techniques to study the effect of dust on the photometry and
spectral energy distributions (SEDs) of galaxies (e.g.\ Bruzual,
Magris \& Calvet 1988; Witt, Thronson \& Capuano 1992; Byun, Freeman
\& Kylafis 1994; Wise \& Silva 1996; Bianchi, Ferrara \& Giovanardi
1996; Corradi, Beckman \& Simonneau 1996; Baes \& Dejonghe 2001a). The
effect of dust attenuation on the observed kinematics, however, is
largely unexplored, and limited to the rotation curves of spiral
galaxies. Davies (1990) argued that the effects of dust absorption on
the observed optical rotation curve of a disc galaxy can lead to a
severe underestimation of the true rotational velocity in the inner
regions. This argument was quantified by Bosma et al.~(1992), who
calculated the effect of dust absorption on the observed rotation
curve of edge-on galaxies. The most detailed work on this subject
comes from Matthews \& Wood (2001), who calculate the effects of dust
attenuation on the rotation curve of low surface brightness (LSB)
galaxies through Monte Carlo simulations.

We have embarked on a program to systematically investigate the
effects of dust on the observed kinematics of galaxies. In previous
papers we focused on the effects of dust attenuation on the observed
stellar kinematics of elliptical galaxies. Initially, we only
considered absorption by dust grains in our models (Baes \&
Dejonghe~2000; Baes, Dejonghe \& De Rijcke~2000). The inclusion of
absorption in the calculation of the observed kinematics is a fairly
simple procedure: it basically adds a weight to the contribution of
each individual star along a line of sight. The radiative transfer
problem becomes much more complicated however when scattering is
also included. In this case, photons can leave their original path,
such that any star can contribute to the observed kinematics in any
line of sight. Moreover, not only the stellar velocities but also the
individual dust grain velocities should be taken into account in the
calculation of the observed kinematics. We argue that the Monte Carlo
method is the only radiative transfer modelling method in which
kinematical information can be included in an elegant and
straightforward way. Using a one-dimensional spherical Monte Carlo
code, we found that the effects of dust scattering are fairly
important: dust attenuation can serve as an additional or alternative
explanation for the stellar kinematical evidence of a dark matter halo
around ellipticals (Baes \& Dejonghe 2001b, 2002a).

In this paper, we tackle the more complicated problem of investigating
the effect of dust attenuation on the observed kinematics of disc
galaxies. We have written a new Monte Carlo code (SKIRT, acronym for
Stellar Kinematics Including Radiative Transfer) that can handle any
geometry of stars and dust. This code is presented in section
{\ref{SKIRT.sec}}, and tested in section {\ref{test.sec}}. We present
our model and the results of our Monte Carlo simulations in section
{\ref{results.sec}}. These results are discussed in section
{\ref{discussion.sec}}, and section {\ref{conclusions.sec}} presents
our conclusions.

\section{Description of the SKIRT code}
\label{SKIRT.sec}

The basic characteristics of Monte Carlo radiative transfer have been
explained at length by various authors (e.g.\ Cashwell \& Everett
1959; Mattila 1970; Witt 1977; Fischer, Henning \& Yorke 1994; Bianchi
et al.\ 1996). In essence, a Monte Carlo radiative transfer code
follows the life of a very large number of individual photons.  A
photon is, at each stage in its existence, characterized by various
quantities, such as its position $\bfr$, propagation direction $\bfk$,
wavelength $\lambda$ and Doppler shift $u$ as measured in a frame of
rest.\footnote{Throughout this paper, we will make no distinction
between the measured Doppler shift $\Delta\lambda$ and the
corresponding line-of-sight velocity $u$, which are related by the
expression $\Delta\lambda/\lambda_0 = u/c$.} After being emitted,
photons propagate on straight lines through the interstellar medium
until they either interact with a dust grain or leave the galaxy. The
various interactions alter the properties of the photon, according to
random numbers generated from the appropriate probability
functions. When at last the photon escapes from the galaxy, its final
properties are recorded. After recording a large numbers of photons in
this way, the global observed properties of the system can be
calculated.

We will not repeat an in-depth description of the principles and the
numerous equations of Monte Carlo radiative transfer here, as they are
well described in the articles listed above. Instead, we will focus on
a number of aspects which make our code different from the existing
ones. These differences are mainly our choice of the dust grid and the
possibility to include kinematical information into the radiative
transfer calculations.

\subsection{The emission process}

When no kinematical information is included in the radiative transfer
calculations, photons are characterized at each moment by a wavelength
$\lambda$, a position $\bfr$ and a direction $\bfk$. Initial values
for these quantities must be generated randomly from the emitting
stellar system, which can be composed of several stellar
components.\footnote{We describe the code as adopted for radiative
transfer calculations in a dusty galaxy. The code can, however,
equally well be applied to any other environment. Terms like ``stellar
component'' must therefore be interpreted in a broad way, and can also
refer to a single star, emitting gas, an AGN, etc., i.e.\ any source
of photons of high enough energy.} The initial emission direction
$\bfk_0$ can be sampled from the unit sphere, the initial position
$\bfr_0$ is sampled from the spatial distribution of the stellar
component, and the initial wavelength $\lambda_0$ is sampled from the
SED of the stellar component at the position $\bfr_0$. The SKIRT code
contains a library with a set of common stellar components (including
spherical Jaffe, Hernquist, Plummer and de Vaucouleurs models, and
axisymmetric exponential, sech, isothermal disc models), from which it
is straightforward to sample a random position (see
Appendix~A). Another library contains a set of SEDs, including Planck
functions and realistic tabulated stellar and galaxy SEDs (Kinney et
al.~1996; Pickles 1998). This library also contains so-called
monochromatic SEDs, which are described by a Dirac delta
function. Sampling a random wavelength from the latter is trivial,
because all photons are emitted with the same wavelength. 

When the SKIRT code is used to calculate the observed kinematics of a
dusty galaxy, photons must also carry with them the kinematic
signature of the star that has emitted them, in the form of a Doppler
shift. Because this is a one-dimensional quantity, not all kinematical
information of the star can be transmitted to the observer, and it is
important to consider how the observed Doppler shift relates to the
velocity vectors of the star and the scattering dust grains. This
problem was discussed in detail in section 2.3 of Baes \& Dejonghe
(2002a), where it is shown that the general expression for the Doppler
shift after $M$ scattering events is
\begin{equation}
	u_M 
	= 
	\bfv_*\cdot\bfk_0 + \sum_{i=1}^M \bfv_{\txd_i}
	\cdot (\bfk_i-\bfk_{i-1}),
\label{vp}
\end{equation}
where $\bfv_*$ and $\bfv_{\txd_i}$ are the velocity vectors of the
emitting star and the $i$th dust grain respectively. The initial value
of the Doppler shift attached to the photon is hence the component of
the star in the direction of the original emission, $u_0 =
\bfv_*\cdot\bfk_0$. 

To complete the initialization of a photon, we also need to generate a
line-of-sight velocity from the appropriate probability
distribution. In this case, this function is the spatial line-of-sight
velocity distribution (LOSVD) $\phi_*(\bfr_0,\bfk_0,u)$, which
describes the probability for a star at a position $\bfr_0$ to have a
velocity component $u$ in the direction $\bfk_0$. In general, the
spatial LOSVD is a marginal probability distribution of the stellar
distribution function $F_*(\bfr,\bfv)$, which describes the
probability density of the stellar component in six-dimensional phase
space,
\begin{equation}
	\phi_*(\bfr,\bfk,u)
	=
	\frac{1}{n_*(\bfr)} 
	\iint F_*(\bfr,\bfv)\, \txd v_{\perp_1}\,\txd v_{\perp_2},
\label{defsLOSVD}
\end{equation}
where $n_*(\bfr)$ is the stellar number density and the integration
covers the entire velocity space perpendicular to the direction
$\bfk$. Thus, in order to be of use for a kinematical Monte Carlo run,
each stellar component must allow the generation of a velocity from
its spatial LOSVDs for arbitrary $\bfr$ and $\bfk$.

The spatial LOSVDs can be calculated analytically for a number of
self-consistent spherical models. Apart from isotropic models
(e.g.~Plummer~1911; H\'enon~1959; Jaffe~1983; Hernquist~1990), this is
possible for the anisotropic families of Plummer and Hernquist models
described by Dejonghe~(1987) and Baes \& Dejonghe~(2002b). For more
complicated dynamical models, however, an analytical evaluation of the
spatial LOSVDs is not possible. For such models, we approximate the
velocity distribution at each position in the galaxy as a local
gaussian distribution. This means that at each position $\bfr$, the
distribution function can be written as
\begin{multline}
	F_*(\bfr,\bfv)
	=
	\frac{n_*(\bfr)}{(2\pi)^{3/2} \sigma_1 \sigma_2 \sigma_3} 
	\\
	\times
	\exp\left[
	-\frac{(v_1-\bar{v}_1)^2}{2\sigma_1^2}
	-\frac{(v_2-\bar{v}_2)^2}{2\sigma_2^2}
	-\frac{(v_3-\bar{v}_3)^2}{2\sigma_3^2}
	\right].
\label{gaussF}
\end{multline}
Here $(\bar{v}_1,\bar{v}_2,\bar{v}_3)$ and
$(\sigma_1,\sigma_2,\sigma_3)$ represent the mean velocities and
velocity dispersions in the directions $(\bfe_1,\bfe_2,\bfe_3)$, the
principle axes of the velocity ellipsoid at the position $\bfr$. The
orientation of the velocity ellipsoid and the values of the mean
velocities and velocity dispersions can vary from position to
position, such that the distribution function (\ref{gaussF})
represents a fairly general distribution function. But a distribution
function of this form has the advantage that the spatial LOSVDs can be
calculated exactly for any direction $\bfk$ by applying the formula
(\ref{defsLOSVD}). We find that the spatial LOSVD will also be a
gaussian distribution,
\begin{equation}
	\phi_*(\bfr,\bfk,u)
	=
	\frac{1}{\sqrt{2\pi}\sigma_u} 
	\exp\left[
	-\frac{(u-\bar{u})^2}{2\sigma_u^2}
	\right],
\label{gaussiansLOSVD}
\end{equation}
with parameters
\begin{gather}
	\bar{u}
	=
	k_1\bar{v}_1 + k_2\bar{v}_2 + k_3\bar{v}_3,
	\\
	\sigma_u^2
	=
	k_1^2\sigma_1^2 + k_2^2\sigma_2^2 + k_3^2\sigma_3^2,
\end{gather}
where obviously $(k_1,k_2,k_3)$ are the components of the vector
$\bfk$ with respect to the orthonormal reference system
$(\bfe_1,\bfe_2,\bfe_3)$. If hence, for each stellar component, we
specify the orientation and parameters of the velocity ellipsoid at
each position $\bfr$, we can easily sample line-of-sight velocities
from the spatial LOSVDs in an arbitrary direction, and hence
initialize the photons' initial Doppler shifts.

\subsection{The dust iteration}

The life of a single photon can be thought of as a loop, whereby at
each iteration (representing a physical process) we must update its
position $\bfr$, propagation direction $\bfk$, wavelength $\lambda$
and Doppler shift $u$. The initial values are determined randomly at
the emission phase, and change at every scattering event. As we only
consider coherent scattering, and we are not taking into account the
re-emission of photons at longer wavelength\footnote{For the present
study, we are primarily interested in the attenuation of starlight by
interstellar dust at optical wavelengths, where the contribution of
re-emitted photons is negligible.}, the only wavelength change of a
photon along its path is due to the varying Doppler shift. These
wavelength variations are so tiny that the optical properties of the
dust (extinction curve, dust albedo and asymmetry parameter) do not
vary. The wavelength of a photon can hence be considered fixed
throughout its lifetime. We must therefore only update the three
quantities $\bfr$, $\bfk$ and $u$ at each iteration, from their old
values ($\bfr_{i-1}\dots$) to the new ones ($\bfr_i\ldots$). This
proceeds in several steps.

The first step in the iteration consists of determining whether an
interaction with a dust grain will take place or whether the photon
will leave the galaxy. To do this, we sample an optical depth
$\tau_\lambda$ from an exponential distribution and compare it to
$\tau_{\text{path},\lambda}$, the total optical depth along the path.
When $\tau_\lambda>\tau_{\text{path},\lambda}$, the photon will leave
the galaxy; in the other case, the photon will interact with a dust
grain. This interaction can either be a scattering or an absorption
event, which is determined by the scattering albedo. In the former
case, the photon will continue its journey through the galaxy, in the
latter case, the loop is ended.

The second step in the loop, if the photon is scattered, is the
determination of the position of the scattering. Therefore we have to
translate the sampled optical depth $\tau_\lambda$ to a physical path
length $s$. With this path length known, the new position is $\bfr_i =
\bfr_{i-1}+s\bfk_{i-1}$.

The third step in the iteration is the determination of the new
propagation direction $\bfk_i$ of the photon. It is found by sampling
a direction from the probability density $p(\bfk_i) =
\Phi_\lambda(\bfk_{i-1},\bfk_i)$, which represents the scattering
phase function. Various phase functions are built into the SKIRT code,
including the isotropic phase function and the anisotropic
Henyey-Greenstein phase function. The sampling of a direction from
these phase functions can be performed analytically.

The final step is to update the photon's Doppler shift. The relative
orientation of the propagation directions before and after the
scattering event cause a change in Doppler shift [from equation
(\ref{vp})]
\begin{equation}
	u_i - u_{i-1}
	= 
	\bfv_{\txd_i}\cdot(\bfk_i-\bfk_{i-1})
\end{equation}
To calculate the updated Doppler shift, we need to sample a dust grain
velocity from the dust velocity distribution. Therefore, the dust
components in the SKIRT library must not only be specified by a
spatial distribution, but also by the entire phase space distribution
$F_\txd(\bfr,\bfv)$. Analogous to the stellar distribution function,
we assume that the dust velocity field can be described by a
trivariate gaussian distribution. Because the dispersions in the
interstellar medium are fairly small compared to the rotational
velocities however, we restrict ourselves to assuming an isotropic
dispersion tensor, such that
\begin{equation}
	F_\txd(\bfr,\bfv)
	=
	\frac{n_\txd(\bfr)}{(2\pi)^{3/2}\sigma_\txd^3}
	\exp\left[
	-\frac{(\bfv-\bar{\bfv})^2}{2\sigma_\txd^2}
	\right],
\end{equation}	
with $\bar{\bfv}$ and $\sigma_\txd$ being the mean velocity vector and
the velocity dispersion of the dust. Similarly as in the stellar case,
we don't have to sample a full three-dimensional velocity vector for
the dust, but only a component in the direction
\begin{equation}
	\bfk_i' 
	= 
	\frac{\bfk_i-\bfk_{i-1}}{||\bfk_i-\bfk_{i-1}||}.
\end{equation}
This means that we just have to sample a velocity from a
one-dimensional gaussian distribution with mean velocity $\bar{u} =
\bar{\bfv}\cdot\bfk_i'$ and dispersion $\sigma_u = \sigma_\txd$.

\subsection{Integration through the dust}
\label{integrdust.sec}

The first step in the iteration described in the previous subsection
requires the calculation of the total optical depth along a given
path. It is found through the integral
\begin{equation}
	\tau_{\text{path},\lambda}(\bfr,\bfk)
	=
	\int_0^\infty
	\kappa_\lambda(\bfr+s\bfk)\,\txd s,
\label{task1}
\end{equation}
where $\kappa_\lambda(\bfr)$ represents the total (absorption plus
scattering) opacity of the dust at a position $\bfr$. To execute the
second step in the iteration, we need to convert an optical depth
$\tau_\lambda$ into a physical path length $s(\bfr,\bfk,\tau_\lambda)$
along a given path, which is done by solving the equation
\begin{equation}
	\int_0^s
	\kappa_\lambda(\bfr+s'\bfk)\,\txd s'
	=
	\tau_\lambda
\label{task2}
\end{equation}
for $s = s(\bfr,\bfk,\tau_\lambda)$. Only for a few simple geometries,
such as constant density or some spherical dust distributions, can
these integrations be performed analytically. In general, the
calculation of the two quantities
$\tau_{\text{path},\lambda}(\bfr,\bfk)$ and
$s(\bfr,\bfk,\tau_\lambda)$ must be performed numerically. These
integrations are usually the most time-consuming part of Monte Carlo
codes. Therefore, it is important to think about an efficient and
flexible way to do this; we have considered two approaches.

\subsubsection{The UDD grid}

In order to allow a completely arbitrary distribution of dust
(including clumpy dust distributions), most modern radiative transfer
Monte Carlo codes adopt an approach that can be described as a uniform
dust density (UDD) grid. It consists of dividing space into a number
of cells, and attaching a uniform dust density to each cell, for
example the value at the centre of the cell. This uniform density (and
hence opacity) makes it very easy to calculate the optical depth
[equation~(\ref{task1})] along a given path: simply calculate the
distance that the photon runs through each cell it crosses, multiply
this distance with the opacity in the cell, and add all the pieces
together. The inversion of the equation (\ref{task2}) is not much more
difficult. First, look for the cell in which the interaction will take
place. Since the opacity increases linearly with path length within a
single cell, this inversion of the equation is then straightforward.

In the SKIRT code, a UDD grid is included based on a three-dimensional
cartesian grid. The grid cells can be chosen arbitrarily, in order to
achieve an efficient grid for a variety of dust distributions. An
algorithm is included to construct clumpy dust distributions from an
underlying smooth dust distribution, as described in Witt \& Gordon
(1996) and Bianchi et al.~(2000).

\subsubsection{The IDD grid}

In addition to the UDD grid, we explore a new approach to handle the
integration through the dust: an interpolated dust density (IDD)
grid. The basis is the same: we divide space into a number of
cartesian grid cells, but, instead of attaching a uniform opacity to
each cell, we use the correct values of the opacity at the eight
border points of the cells, and we apply a simple trilinear
interpolation routine to determine the opacity in each of the points
within a grid cell. Consider a photon at the position $\bfr=(x,y,z)$,
a position within the $(i,j,k)$'th cell with border points $x_i\leq
x\leq x_{i+1}$, etc.  We define the dimensionless quantities $p$, $q$
and $r$ as
\begin{equation}
	p = \frac{x-x_i}{x_{i+1}-x_i}, \quad
	q = \frac{y-y_j}{y_{j+1}-y_j}, \quad
	r = \frac{z-z_k}{z_{k+1}-z_k}.
\end{equation}
Since $p$ increases linearly from 0 to 1 as $x$ moves linearly from
$x_i$ to $x_{i+1}$, and similarly for $q$ and $r$, the trilinear
interpolation rule for the opacity simply reads
\begin{align}
	\kappa_\lambda(\bfr)
	&=
	\left(1-p\right)
	\left(1-q\right)
	\left(1-r\right)
	\kappa_{i,j,k}
	\nonumber \\
	&\quad+
	\left(1-p\right)
	\left(1-q\right)
	r\,
	\kappa_{i,j,k+1}
	\nonumber \\
	&\quad+
	\cdots
	\nonumber \\
	&\quad+
	p\,q\,r\,\kappa_{i+1,j+1,k+1},
\label{kappaintpol}
\end{align}
where $\kappa_{i,j,k}$, etc.\ represent the opacity in the border
points. Now consider the fraction of the path that remains in this
cell. This path can be parametrized using the expression $\bfr+s\bfk$,
with the path length $s$ as a free parameter. In each of the points
along this path (as long as we remain in the same cell), we can
calculate the opacity by a similar formula as (\ref{kappaintpol}),
\begin{align}
	\kappa_\lambda(\bfr+s \bfk)
	&=
	\left[1-p(s)\right]
	\left[1-q(s)\right]
	\left[1-r(s)\right]
	\kappa_{i,j,k}
	\nonumber \\
	&\quad+
	\left[1-p(s)\right]
	\left[1-q(s)\right]
	r(s)\,
	\kappa_{i,j,k+1}
	\nonumber \\
	&\quad+
	\cdots
	\nonumber \\
	&\quad+
	p(s)\,q(s)\,r(s)\,\kappa_{i+1,j+1,k+1},
\end{align}
where the parameters $p$, $q$ and $r$ are now linear functions of the
path length $s$,
\begin{gather}
	p(s) = \frac{x+s\,k_x-x_i}{x_{i+1}-x_i}, \\
	q(s) = \frac{y+s\,k_y-y_j}{y_{j+1}-y_j}, \\
	r(s) = \frac{z+s\,k_z-z_k}{z_{k+1}-z_k}.
\end{gather}
Along a path within a single cell, the opacity is thus a cubic
polynomial in $s$, the distance traveled along the path,
\begin{equation}
	\kappa_\lambda(\bfr+s \bfk)
	=
	\sum_{m=0}^3 a_m\,s^m,
\end{equation}
with constant coefficients $a_m$. With this expression for the
opacity, the calculation of the optical depth
$\tau_{\text{path},\lambda}(\bfr,\bfk)$ is very
straightforward. Indeed, it is sufficient to calculate, for each cell
the path passes through, the two boundary positions. This enables us
to calculate the path length $s$ within the cell and the coefficients
$a_m$. Given these coefficients, the portion of the optical depth in
that cell is found by substitution into the quartic polynomial
\begin{equation}
	\int_0^s \kappa_\lambda(\bfr+s'\bfk) \,\txd s'
	=
	\sum_{m=1}^4 \frac{a_m\,s^m}{m}.
\end{equation}
Also, the calculation of the path length $s(\bfr,\bfk,\tau_\lambda)$ is
straightforward. First, we determine the cell in which the optical
depths at the border points bracket the value of $\tau_\lambda$, in a
similar way as described above to calculate the total optical depth
along the path. The determination of $s$ then comes down to finding
the root of a quartic equation. The obvious way of finding this root
is by means of a Newton-Raphson iteration, because we can easily
compute both the function values and the derivatives of this
function. This routine has a very fast quadratic convergence (the
number of significant digits approximately doubles with each
iteration). Usually, no more than two iterations are necessary.

\subsubsection{Comparison of the grids}

\begin{figure}
\centering
\includegraphics[clip, bb=103 82 446 550, angle=-90, width=0.45\textwidth]{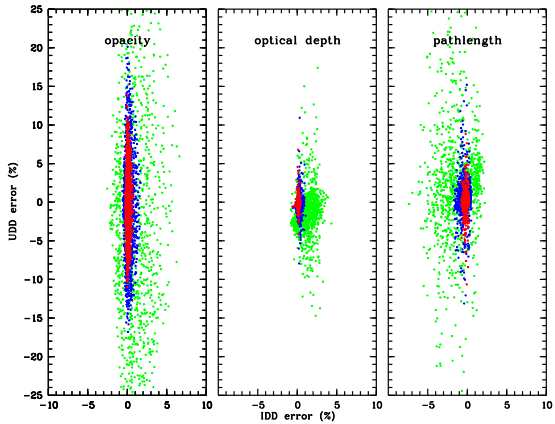}
\caption{ A comparison of the accuracy of the integrations through a
UDD dust grid (constant density) and the IDD dust grid (trilinear
interpolation). The three panels contain the relative errors of the
opacity $\kappa_\lambda(\bfr)$, the optical depth along a path
$\tau_{\text{path},\lambda}(\bfr,\bfk)$ and the path length
$s(\bfr,\bfk,\tau_\lambda)$ for 1000 random photons. The dust grids in
both cases contain the same cells. The different colors correspond to
dust grids with 100$^3$ (green), 200$^3$ (blue) and 300$^3$ (red)
cells respectively.}
\label{grid.eps}
\end{figure}

To compare the accuracy and efficiency of the two approaches, we
applied both techniques to a simple spherical dust geometry,
characterized by an opacity
\begin{equation}
	\kappa_\lambda(\bfr)
	\propto
	\left(1+\frac{r^2}{c^2}\right)^{-\alpha/2}.
\end{equation}
The integrations through the dust, i.e.\ the calculation of the two
functions $\tau_{\text{path},\lambda}(\bfr,\bfk)$ and
$s(\bfr,\bfk,\tau_\lambda)$, can be performed exactly for such a dust
component with $\alpha=2$, 3 or 5 [see Baes (2001) for details]. These
exact results were compared with the results from the UDD and IDD
grids. In figure~{\ref{grid.eps}} we plot the relative error of the
calculated values of the opacity, the total optical depth along a path
and the path length corresponding to a given optical depth for 1000
randomly generated photons for such a model. For a given number of
dust grid cells, the accuracy of the integrations in the IDD grid is
much better than those in the UDD grid. However, in terms of
computational efficiency, comparing the two grids with the same
numbers of cells is not fair, because the integrations in the UDD grid
are less complicated and hence faster. We found that the IDD grid
integrations are a factor 3 slower in the mean. On the other hand, in
order to reach a similar accuracy as the IDD grid, many more cells are
necessary in the UDD grid. Apart from slowing down the integrations,
this will also increase the memory requirements significantly. After
several tests, we found that the IDD grid is computationally
preferable above the UDD grid, in particular when the system harbours
a large dynamical range of dust densities (such as an exponential
profile which is appropriate in disc galaxies). It should be noted
however, that the UDD grid still remains a very useful alternative, in
particular to include clumpy dust distributions into the SKIRT code.

\subsection{Optimization of the code}
\label{opt.sec}

The procedure described at the beginning of the previous section is
the most basic version of a Monte Carlo iteration step. When executed
as such, however, it would be very inefficient. Throughout the years a
number of 'intelligent tricks' have been presented which increase the
computational efficiency considerably. Most of them attach a weight to
each photon, which can be altered during its lifetime. We incorporate
three such techniques into our SKIRT code.
 
The first optimization to the basic Monte Carlo scenario is to turn
all interactions into scattering events. In reality, of course, the
interaction between a photon and a dust grain can either be an
absorption or a scattering event; the nature of this event is
determined by the albedo $\omega_\lambda$ of the dust grains. The
problem with absorptions is that the photon is lost, and therefore
does not contribute to the observed radiation field anymore. To
overcome this problem, we force each interaction to be a scattering
event and alter the weight of the photon after each scattering by a
factor $\omega_\lambda$ in order to compensate for the fraction of
absorbed photons.

The second technique is the peeling-off procedure introduced by
Yusef-Zadeh et al.\ (1984). In the basic Monte Carlo iteration, each
photon leaves the galaxy at a certain stage, and it can be classified
in bins according to its position and direction. This approach has a
number of disadvantages. Firstly, if we are interested in only one
single observing position, most of the photons are not used. Secondly,
if the observer has the misfortunate of being located in a direction
in which not many photons leave the galaxy, he might have to wait for
a very long time to detect enough photons to obtain good
statistics. Thirdly, if one wants to compare the view of a system from
two observing positions close to each other, the direction bins must
be chosen to be very small, which requires a very large number of
photons. The peeling-off technique can be used to tackle these
problems. It consists of creating a new photon during each scattering
[resp.\ emission] process, which is scattered [resp.\ emitted] exactly
in the direction of the observer. This photon will be detected by the
observer. Its weight is altered by a factor
$\Phi(\bfk,\bfk_{\text{obs}})$ [resp.\ 1] to compensate for the
probability of scattering [resp.\ emission] in the direction
$\bfk_{\text{obs}}$, and a factor
$\exp[-\tau_{\text{path},\lambda}(\bfr,\bfk_{\text{obs}})]$ to
compensate for extinction along the line-of-sight.

The third optimization we apply is the principle of forced first
scattering (Witt 1977), which is useful in the relatively low optical
depth regime. A problem in such systems is that most of the photons
leave the galaxy directly without a single interaction with a dust
grain, and therefore many photons need to be generated in order to
have a good statistics of the scattered radiation. To overcome this
problem, we split the emitted photon into two parts. The first part of
the photon directly leaves the galaxy and obtains a weight
$\exp(-\tau_{\text{path},\lambda})$ (because we adopt the peeling-off
technique, we don't have to consider this fraction anymore). The other
part of the photon with the remaining weight is forced to be scattered
at least once, which is achieved by sampling an optical depth
$\tau_\lambda$ such that
$\tau_\lambda<\tau_{\text{path},\lambda}$. This technique is called
the principle of forced first scattering, but nothing prevents us
applying it only to freshly emitted photons: it can equally well be
applied to scattered photons (Bianchi et al.~1996). The number of
forced scatterings we apply in our calculations is usually 3, but this
value can be set arbitrarily according to the studied system.

\subsection{The detection and data reduction processes}

The last phase in the Monte Carlo cycle is the detection phase,
whereby the information contained in the photon is transmitted to the
observer. Because of the peeling-off procedure, the number and
position of the observers can be chosen randomly. At each observing
position, the user can choose from several simulated instruments to
detect the photons: photometers, low-resolution and high-resolution
spectrographs. Each instrument consists of a number of pixels on the
plane of the sky and they come in two geometries: rectangular and
circular. Each pixel contains a detector, the nature of which depends
on the kind of instrument.

The photometers are the most simple instruments, whereby the detector
in each pixel is just a counter. Every time a photon hits the pixel
and its wavelength is within the detector's bandwidth, its weight is
added to the intensity. At the end of the simulation, the 2D surface
brightness image of the system is directly obtained. The photometers
are designed such that each photon remembers the number of
scatterings it has undergone, and this information is also used by the
photometric detectors. This allows us to construct images for each
individual scattering (direct emission, photons that have been
scattered once, etc.), which can be useful to disentangle the effects
of absorption and scattering. 

The low-resolution spectrographs work in a very similar way, except
that there is not just one detector, but a range of detectors
corresponding to different wavelength bins. In this way a data cube is
produced (actually a set of data cubes, one for every scattering),
very similar to the data cubes obtained by radio or Fabry-Perot
observations. The high-resolution spectrographs basically produce
similar data cubes, with Doppler shift bins replacing the wavelength
bins. Combining the different Doppler shift bins at a given position
$\bfx$ on the sky, we obtain the LOSVD or line profile
$\phi_\txp(\bfx,u)$, which describes the entire projected velocity
distribution. From the LOSVDs, the 2D mean projected velocity field
$v_\txp(\bfx)$ and projected velocity dispersion field $\sigma_\txp(\bfx)$
can be calculated.

From a conceptual point of view, high-resolution and low-resolution
spectrographs work in a very similar way. Both separate the incoming
photons in different bins according to the extra information carried
by the photon, which is the wavelength in the former and the Doppler
shift in the latter case. From a computational point of view, however,
there is an important difference between the two. Indeed, the
wavelength of the photon is actively used throughout the Monte Carlo
iteration, because the optical properties of the dust grains depend on
the photon's wavelength. On the contrary, a photon's path does not
depend on its Doppler shift. Moreover, the changes in Doppler shift
experienced by a photon when it moves through the dust are independent
of the initial stellar Doppler shift. Therefore, we can optimize our
code in the following way. Instead of initializing the photon's
Doppler shift by generating a $u_0$ from the stellar spatial LOSVD, we
postpone this contribution of the stellar velocity until later and
initialize $u_0$ to zero. When the photon leaves the galaxy after $M$
scattering events, it will then carry a Doppler shift
\begin{equation}
	u_M^\txd
	=
	\sum_{i=1}^M \bfv_{\txd_i}
	\cdot (\bfk_i-\bfk_{i-1}),
\end{equation}
where we still have to add the stellar Doppler shift in order to
obtain the correct value. Instead of generating a single initial
stellar Doppler shift $u_0$ from the spatial LOSVD
$\phi_*(\bfr_0,\bfk_0,u)$, we split the photon over all velocity
bins. Each fraction will then be weighted by a weight factor that
represents the probability that the final Doppler shift of the photon
will fall within the corresponding bin. In order to fall within a bin
with boundaries $u_k$ and $u_{k+1}$, the photon must have an initial
stellar Doppler shift satisfying $u_k \leq u_0+u_M^\txd \leq
u_{k+1}$. The weight factor corresponding to the $k$th bin will hence
equal
\begin{equation}
	w_k 
	= 
	\int_{u_k-u_M^\txd}^{u_{k+1}-u_M^\txd} 
	\phi_*(\bfr_0,\bfk_0,u)\,
	\txd u.
\label{wk}
\end{equation}
When we substitute the gaussian spatial LOSVD (\ref{gaussiansLOSVD})
into this expression, we can evaluate this weight factor exactly
\begin{equation}
	w_k 
	= 
	\frac{1}{2}
	\left[
	\erf\left(	
	\frac{u_{k+1}-u_M^\txd-\bar{u}}
	{\sqrt{2}\sigma_u}
	\right)
	-
	\erf\left(	
	\frac{u_k-u_M^\txd-\bar{u}}
	{\sqrt{2}\sigma_u}
	\right)
	\right].
\end{equation}
This method is of the same kind as the optimization techniques
described in section~{\ref{opt.sec}} and also increases the efficiency
of the SKIRT code in a significant way.

\subsection{General architecture and performance}

The SKIRT code is implemented in the ANSI standard C++ language and
makes optimal use of the object-oriented structure of this
language. The output of the code consists of FITS files, which are
created by means of the FITSIO and CCFITS packages. 

The run time of the SKIRT code with reliable output depends critically
on the number of photons, the number of observing directions, the
spatial and velocity resolution of the instruments and the number of
cells in the dust grid. For the simulations described in
section~{\ref{results.sec}}, we found that about $5\times10^7$ photons
are necessary to obtain accurate results. The code spends 60 per cent
of the time moving through the dust grid, while most of the remaining
time is spent in the calculation of the spatial LOSVDs. When run on a
modern Pentium {\sc iv} PC, the run time of a single simulation (with
7 observing positions) is of the order of 50 to 100 hours.

\section{Testing the code}
\label{test.sec}

The code was first run without dust included in order to test the
projection algorithm and the calculation of the observed
kinematics. We used simple non-rotating spherical geometries which
allow an analytical expression of the surface brightness and projected
velocity dispersion profiles (Dejonghe~1987; Hernquist~1990;
Jaffe~1983, Baes \& Dejonghe~2002b). Also the surface brightness
profiles of face-on and edge-on exponential discs (which can be
calculated analytically) are accurately recovered. Moreover, when we
attach a simple velocity structure to an exponential disc, with a flat
rotation curve and exponentially decreasing velocity dispersions, the
observed kinematics can be also calculated analytically in the face-on
and edge-on directions. These are recovered very well, proving the
correct behaviour of the code.

Next, we compared the results of the SKIRT code with our previous
Monte Carlo calculations of the observed kinematics of elliptical
galaxies (Baes \& Dejonghe 2001b, 2002a). The Monte Carlo code used in
these papers also allowed for the calculation of the observed
kinematics, but only for spherical models, chosen such that the
integrations through the dust could be performed analytically. We found
an excellent agreement between these results and those calculated with
the new SKIRT code.

\begin{figure}
\centering
\includegraphics[clip, bb=137 285 461 741,width=0.4\textwidth]{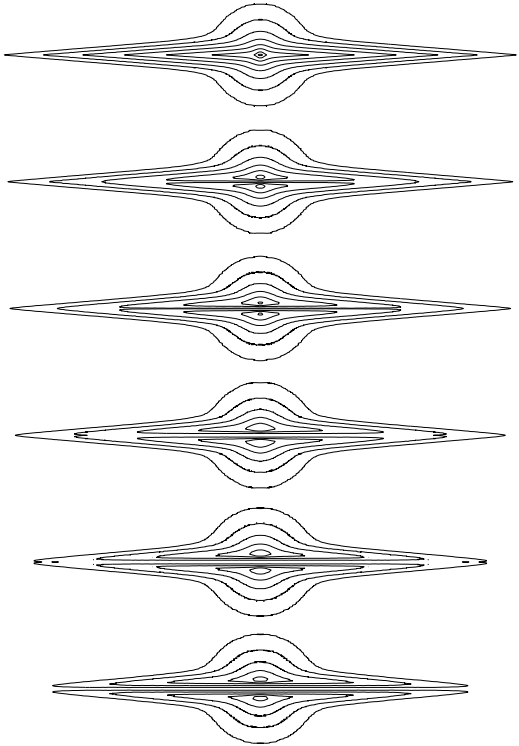}
\caption{ The $B$-band surface brightness distribution of an edge-on
disc galaxy, for various values of the optical depth. The model is
similar to the BT0.3 model described by Byun et al.~(1994), i.e.\ an
exponential stellar disc with scale length $h_*=4$~kpc and scale
height $z_*=0.35$~kpc, a de Vaucouleurs bulge with $R_{\text{e}} =
1.6$~kpc and a bulge-to-total luminosity ratio of 0.3. The dust is
also exponential with scale length $h_\txd=4$~kpc and
$z_\txd=0.14$~kpc, and the face-on central optical depths in the
different plots are $\tau_\txV=0$, 0.5, 1, 2, 5 and 10 (from top to
bottom). Surface brightness contours are shown with a step of
$\Delta\mu=1$~mag/kpc$^2$. }
\label{Byun.eps}
\end{figure}

Finally, we compared our (photometric) results with the radiative
transfer calculations of other authors. In figure~{\ref{Byun.eps}} we
plot the two-dimensional edge-on surface brightness distribution of an
Sb galaxy model in the $B$ band for various values of the optical
depth. The models are the same models as the BT0.3 models of Byun et
al.~(1994), and they consist of a stellar disc, a stellar bulge and a
dust disc. Comparing this figure with their figure~4b, we find an
excellent agreement between the two results. Notice that Byun et
al.~(1994) adopted a completely different approach to solve the
radiative transfer problem.

\section{The observed kinematics of dusty disc galaxies}
\label{results.sec}

We apply the SKIRT code to investigate the effect of dust absorption
and scattering on the photometry and the stellar kinematics of normal
galactic discs. We limit our modelling only to a stellar disc and do
not include a bulge, because we are primarily interested in the disc
kinematics.

\subsection{Presentation of the model}

Our model consists of a simple axisymmetric double exponential disc
galaxy, with stellar emissivity
\begin{equation}
	\ell_\lambda(\bfr)
	=
	\frac{L_\lambda}{4\pi h_*^2 z_*}
	\exp\left(-\frac{R}{h_*}\right)
	\exp\left(-\frac{|z|}{z_*}\right).
\end{equation}
The luminosity $L_\lambda$ is a free scaling parameter in our models,
as both the input emissivity and the output light profile scale with
the luminosity.  For the scale length and scale height we chose the
values adopted by Byun et al.~(1994) to describe the Milky Way,
$h_*=4$ kpc and $z_*=350$ pc. As we focus on the observed kinematics,
which are always measured in a very narrow wavelength region, we chose
a trivial monochromatic SED centered on the $V$ band.

In order to run a kinematical simulation for such a disc galaxy, we
also need to provide details on the kinematical structure of the
disc. We assume that the stellar velocities at each position in the
disc can be represented by a trivariate Gaussian distribution, 
\begin{equation}
	F_*(\bfr,\bfv) 
	\propto 
	\exp \left[ 
	-\frac{v_R^2}{2\sigma_R^2} 
	-\frac{(v_\varphi-\bar{v}_\varphi)^2}{2\sigma_\varphi^2}
	-\frac{v_z^2}{2\sigma_z^2} 
	\right].
\end{equation}
Such a velocity distribution was first proposed by Schwarzschild
(1907) to describe the distribution of stellar velocities in the solar
neighbourhood. The mean azimuthal velocity and the three velocity
dispersion components can vary from position to position. The entire
kinematical structure of the disc will be completely determined if we
have expressions for the four functions $\bar{v}_\varphi(\bfr)$,
$\sigma_R(\bfr)$, $\sigma_\varphi(\bfr)$ and $\sigma_z(\bfr)$. Since
the system is axisymmetric, these quantities will depend only on the
cylindrical radius $R$ and the height $z$. Moreover, because stellar
discs are thin and the contribution of photons high above or under the
symmetry plane of the galaxy will be quite small due to the vertical
exponential decrease, we will assume that these four functions are a
function of $R$ only.

For the $\bar{v}_\varphi$ profile, we assume the arctan profile, a simple
two-parameter fitting function that fits the observed rotation curves
of disc galaxies fairly well (e.g.\ Courteau 1997),
\begin{equation}
	\bar{v}_\varphi(\bfr) 
	=
	\frac{2v_{\text{max}}}{\pi} 
	\arctan \left( \frac{R}{R_0} \right).
\end{equation}
We have adopted an amplitude $v_{\text{max}}=220$ km/s and a scale
radius $R_0=0.5$ kpc, which results in a rotation curve that has a
steep initial gradient and reaches its flat part fairly quickly. The
internal velocity dispersion profiles in disc galaxies are not well
constrained. For a thin self-gravitating disc, the Jeans equations
yield that the vertical velocity dispersion scales with the square
root of the surface density. If we assume a constant mass-to-light
ratio, we obtain the following parametrization,
\begin{equation}
	\sigma_z(\bfr) 
	=
	\sigma_{z0}
	\exp\left(-\frac{R}{2h_*}\right).	
\end{equation}
As all dispersions in a stellar disc are thought to exist from the
gradual stellar heating of an initial cold disc, it makes sense to
assume that the velocity dispersions in the other direction are
intimately coupled to the vertical dispersion. Therefore, we assume a
constant axis ratio of the velocity ellipsoid in our model, such that
$\sigma_R$ and $\sigma_\varphi$ have the same spatial variation as
$\sigma_z$. Notice that such behaviour is found to be in agreement
with the sparse existing data on the velocity dispersions in external
galaxies (Gerssen et al.\ 1997, 2000). The only remaining quantities
that need to be set are normalizations of the dispersions. We adopted
the values $(\sigma_{R0} ,\sigma_{\varphi0} ,\sigma_{z0}) =
(100,75,50)$ km/s. At a galactocentric radius of 8 kpc, this gives
velocity dispersions of (37,28,18) km/s, which roughly corresponds to
the measured velocity dispersions in the solar neighbourhood.

The dust was chosen to be distributed in a similar way as the stellar
distribution,
\begin{equation}
	\kappa_\lambda(\bfr)
	=
	\frac{\tau_\lambda}{2z_\txd}
	\exp\left(-\frac{R}{h_\txd}\right)
	\exp\left(-\frac{|z|}{z_\txd}\right),
\end{equation}
whereby $\tau_\lambda$ (here) represents the face-on optical depth
through the centre of the galaxy,
\begin{equation}
	\tau_\lambda
	=
	\int_{-\infty}^\infty \kappa_\lambda(\bfr)\,\txd z.
\end{equation}
For the dust scale length and scale height, we again choose the values
from Byun et al.~(1994), $h_\txd=4$ kpc and $z_\txd=140$ pc. Notice that the
dust is hence significantly more confined to the plane of the galaxy,
a feature which gives rise to the well-known dust lanes in edge-on
galaxies. For the velocity field of the dust grains, we assumed an
isotropic gaussian distribution,
\begin{equation}
	F_\txd(\bfr,\bfv) 
	\propto 
	\exp \left[ 
	-\frac{v_R^2+(v_\varphi-\bar{v}_\varphi)^2+v_z^2}{2\sigma_\txd^2} 
	\right].
\end{equation}
We assumed the same mean velocity as the stellar distribution, and a
velocity dispersion $\sigma_\txd=8$~km/s, which is a typical value for
the dispersion of the cold gas in the interstellar medium (Sofue \&
Rubin 2001). Finally, for the optical properties of the dust we took
the average Galactic values (Gordon, Calzetti \& Witt~1997).

These functions and parameters completely determine our model, except
for the one free parameter $\tau_\txV$, the central face-on optical
depth in the $V$ band (the corresponding optical depth at other
wavelengths is determined by the assumed optical properties of the
dust). As discussed in the Introduction, the optical depth in spiral
galaxies is still a matter of debate. We ran our models for the
optical depth values $\tau_\txV=0$, 0.5, 1, 2, 5 and 10, in order to
cover a wide range of possible scenarios.

\subsection{Modelling results}

\begin{figure*}
\centering
\includegraphics[clip, bb=78 50 400 456, width=58mm]{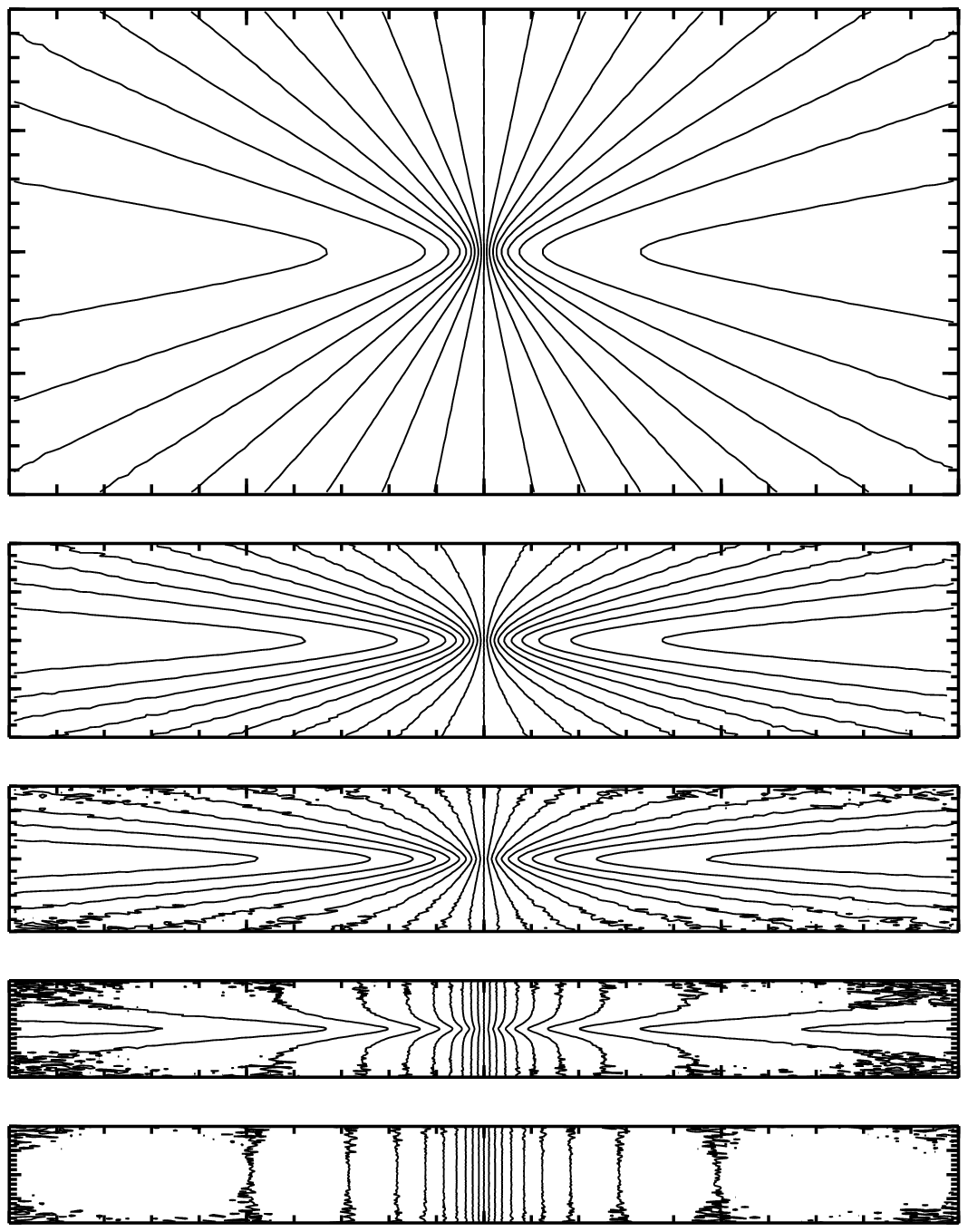}
\includegraphics[clip, bb=78 50 400 456, width=58mm]{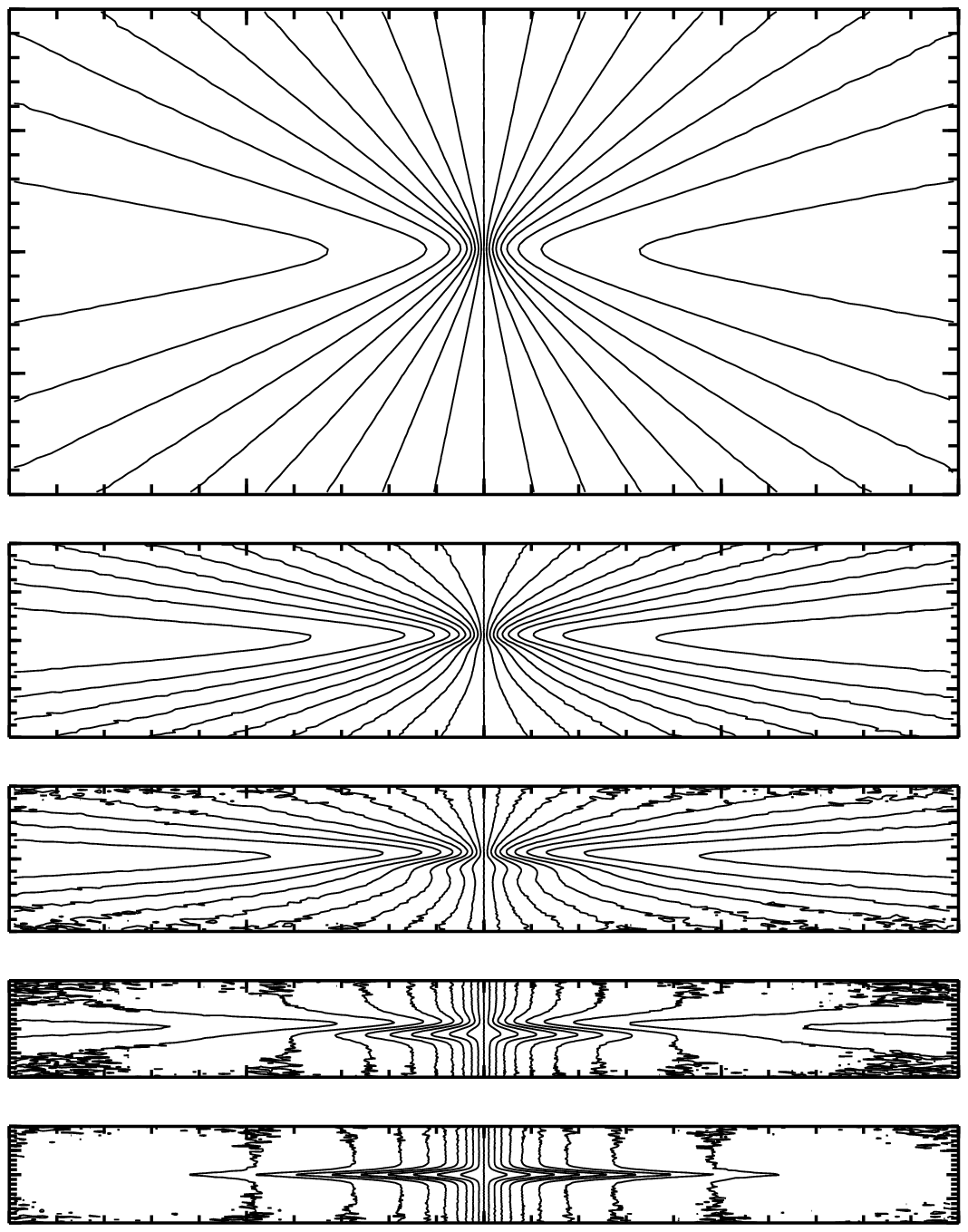}
\includegraphics[clip, bb=78 50 400 456, width=58mm]{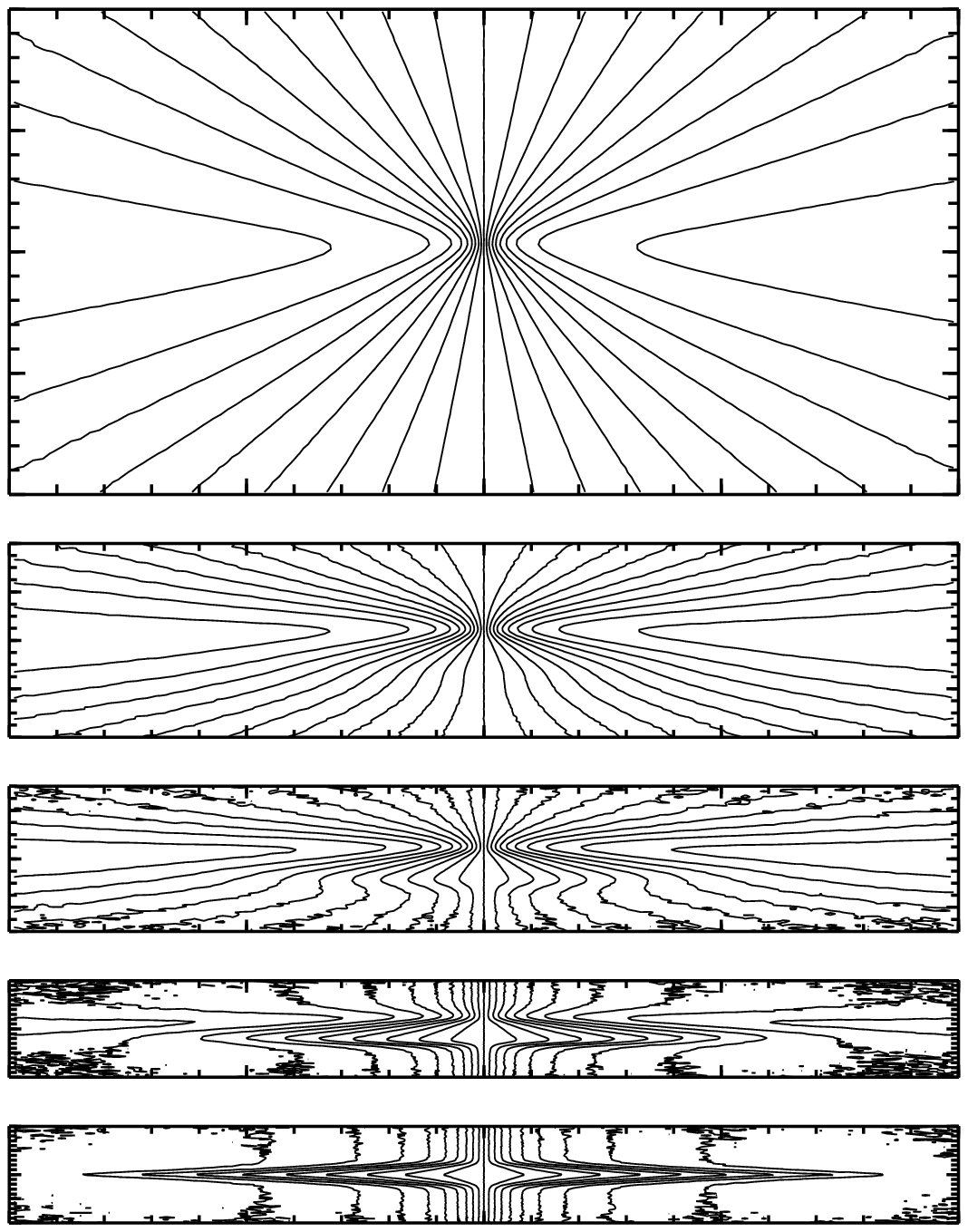}
\caption{ The effect of dust attenuation on the projected mean
velocity field of our galaxy disc models. Each single panel is a
diagram with contours of equal projected mean velocity (the so-called
spider diagram). The total field of view along the major axis is 40
kpc, i.e.\ 5 scale lengths at each side of the galaxy centre. The
different contours in the panels are drawn with a step $\Delta
v_\txp=20$~km/s. The five rows correspond to different inclination
angles, ranging from $i=60$ (top row) over $i=80, 85, 88$ to the
edge-on $i=90$ (bottom row). The three different columns correspond to
different values of the optical depth: the left column is the
dust-free model, whereas the middle and right ones correspond to
$\tau_\txV=1$ and $\tau_\txV=5$ respectively. }
\label{MW_vp.eps}
\end{figure*}

\begin{figure*}
\centering
\includegraphics[clip, bb=78 50 400 780, width=58mm]{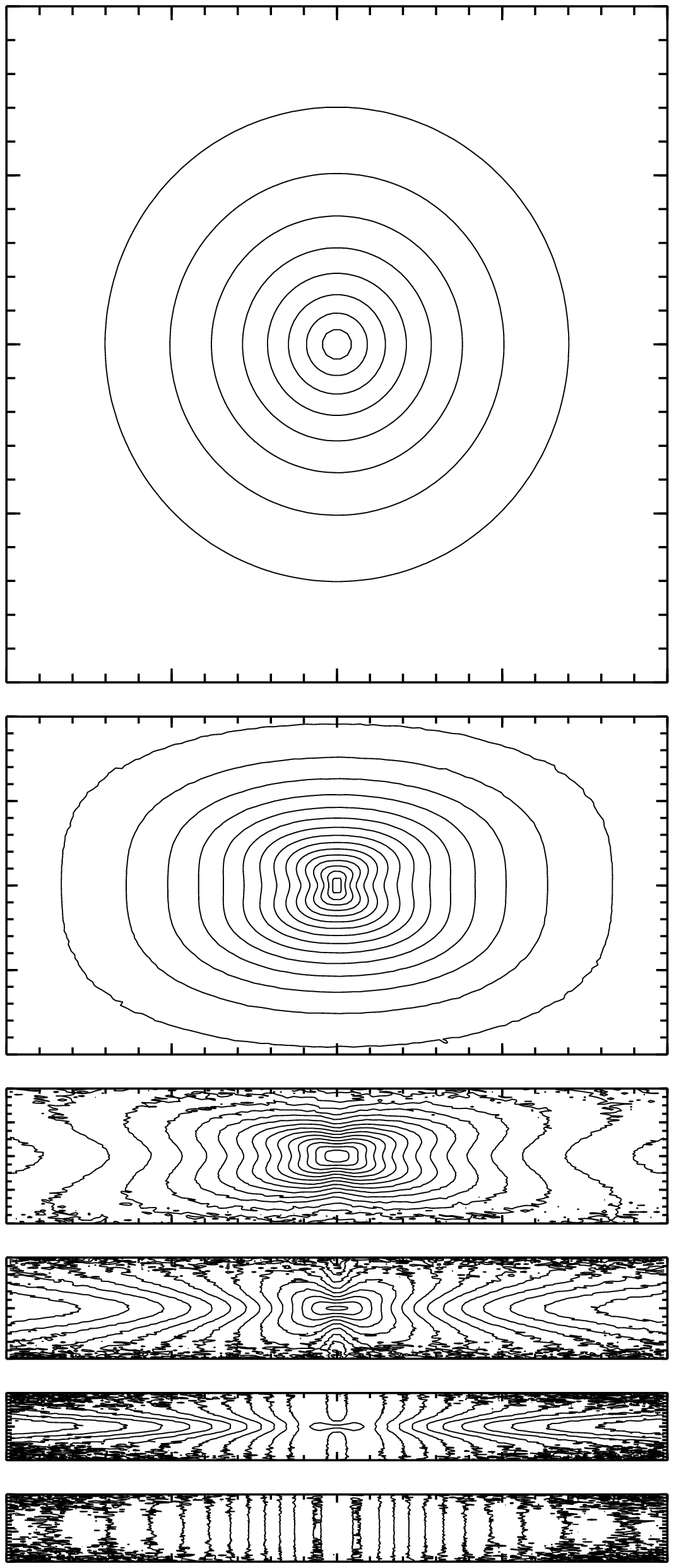}
\includegraphics[clip, bb=78 50 400 780, width=58mm]{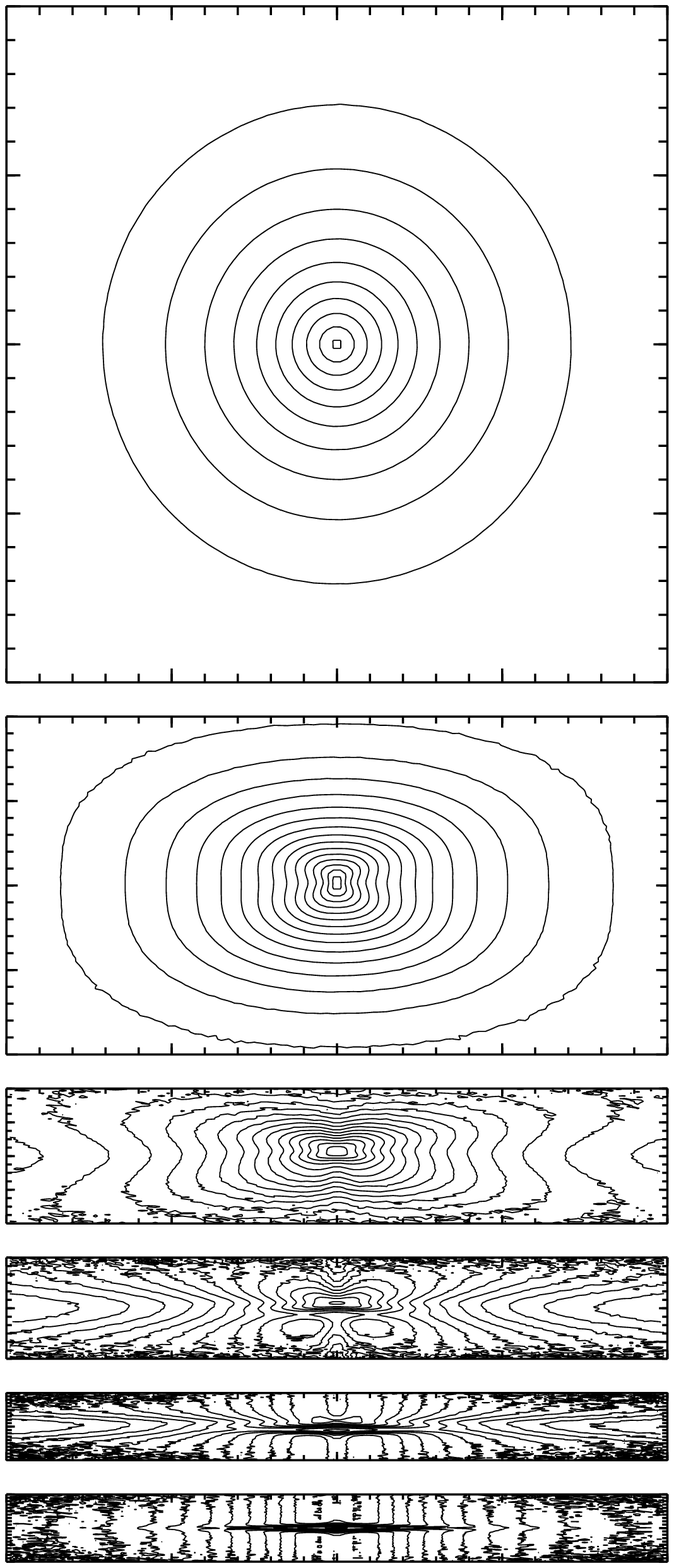}
\includegraphics[clip, bb=78 50 400 780, width=58mm]{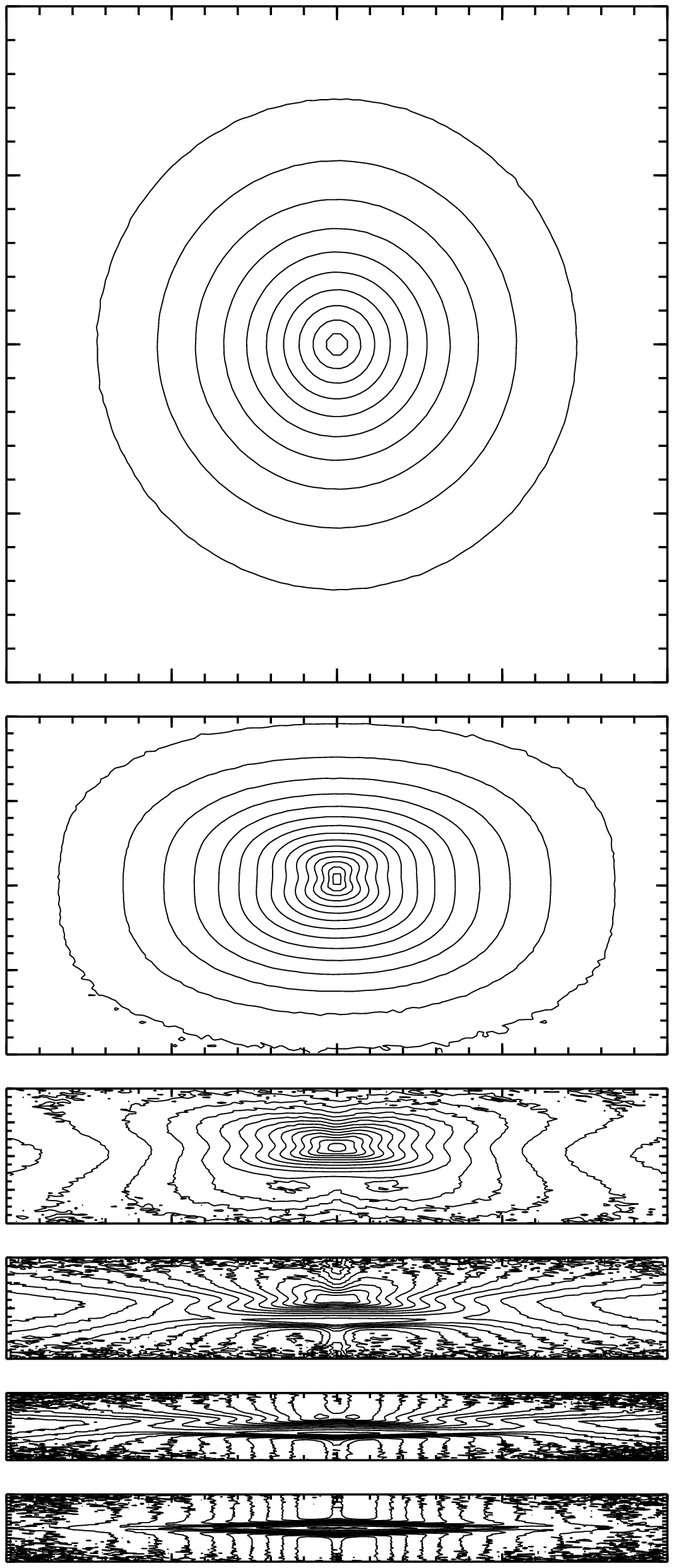}
\caption{ The effect of dust attenuation on the projected velocity
dispersion field of our galaxy disc models. Each single panel is a
diagram with contours of equal projected velocity dispersion. The
different contours are drawn with a step $\Delta \sigma_\txp=5$~km/s.
The layout of this figure is similar to figure~{\ref{MW_vp.eps}},
except that an extra panel at the top is added representing the
face-on direction. }
\label{MW_sigp.eps}
\end{figure*}

\begin{figure*}
\centering
\includegraphics[clip, bb=43 148 547 493, width=\textwidth]{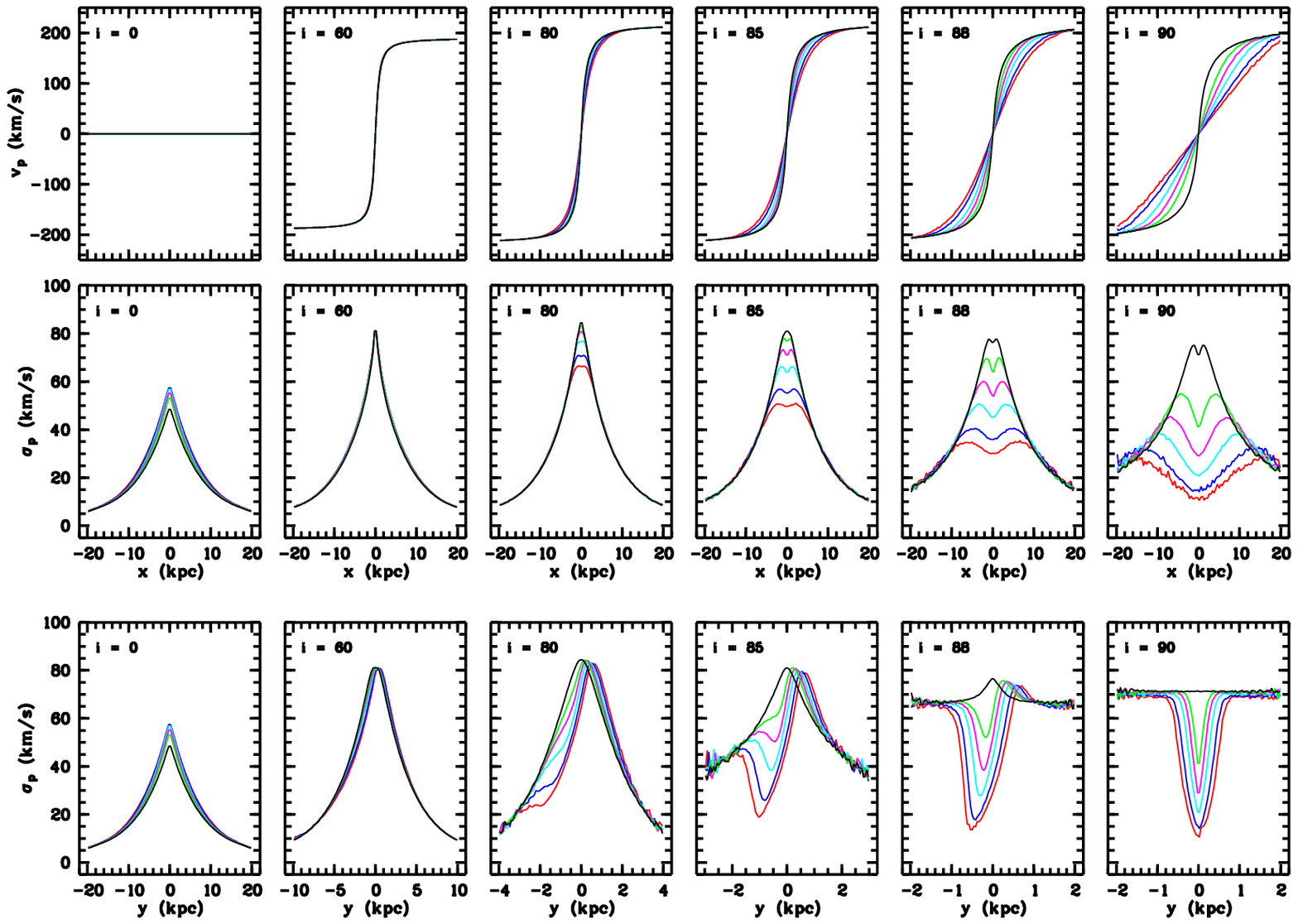}
\caption{ The observed kinematics of our galaxy disc models along the
major and minor axes. The panels on the two rows represent the mean
projected velocity profiles and projected velocity dispersion profiles
along the major axis, and the bottom panel show the minor axis
projected velocity dispersion profiles. The different columns
correspond to different inclination angles, ranging from face-on
(left) to edge-on (right). Notice that the projected velocity profiles
are the observed profiles, and are not corrected for inclination. The
different curves in each panel correspond to different values of the
optical depth: $\tau_\txV=0$ (black), $\tau_\txV=0.5$ (green), $\tau_\txV=1$
(magenta), $\tau_\txV=2$ (cyan), $\tau_\txV=5$ (blue) and $\tau_\txV=10$
(red).}
\label{MWkin.eps}
\end{figure*}

In this section we present and describe the results of our Monte Carlo
simulations. We will not discuss the effects of dust on the
photometry, since such results have been discussed at length by other
authors, such as Byun et al.~(1994), Corradi et al.~(1996) and Bianchi
et al.~(1996). Instead we concentrate on the effects of dust on the
observed kinematics, in particular on the mean projected velocity
$v_\txp(\bfx)$ and the projected velocity dispersion
$\sigma_\txp(\bfx)$.

In figures~{\ref{MW_vp.eps}} and~{\ref{MW_sigp.eps}} we compare, at
various inclination angles, the 2D mean projected velocity and
projected velocity dispersion fields of an optically thin galaxy
(left) with the corresponding fields of a dusty galaxy with optical
depth $\tau_\txV=1$ (centre) and $\tau_\txV=5$ (right)
respectively. Notice that the mean projected velocity field is not
plotted in the face-on direction, as there is no rotation
perpendicular to the galactic plane. To facilitate the interpretation
of these 2D kinematical fields, we plotted in figure~{\ref{MWkin.eps}}
the major and minor axis kinematical profiles of our models at various
inclination angles, explicitly as a function of the optical depth.

\subsubsection{Edge-on galaxies}

\begin{table}
\centering
\caption{The effect of dust attenuation on the slope of the mean
projected velocity curve. This slope is expressed through the major
axis radius at which the rotation curve reaches half of the asymptotic
value $v_\txp^{\text{max}} = 220 \sin i$~km\,s$^{-1}$. These radii are
tabulated for the various inclinations and optical depths considered
in our simulations. For a infinitely thin and dust-free galaxy, the
intrinsic value would be 0.5\,kpc, independent of the inclination
angle.}
\begin{tabular}{lrrrrr}
\hline
$\tau_\txV$ & $i=90$ & $i=88$ & $i=85$ & $i=80$ & $i=60$ \\ \hline
0 & 2.05 & 1.58 & 1.24 & 0.96 & 0.66 \\
0.5 & 4.25 & 2.22 & 1.35 & 0.95 & 0.66 \\
1 & 5.69 & 2.93 & 1.69 & 1.04 & 0.65 \\
2 & 7.26 & 3.71 & 2.10 & 1.24 & 0.65 \\
5 & 9.23 & 4.77 & 2.72 & 1.56 & 0.70 \\
10 & 10.63 & 5.58 & 3.21 & 1.82 & 0.75 \\ \hline \\
\end{tabular}
\label{rot.tab}
\end{table}

For edge-on galaxies, the effects of dust on both the mean projected
velocity and the projected velocity dispersion are very strong, as is
obvious from the bottom row panels of figure~{\ref{MW_vp.eps}}
and~{\ref{MW_sigp.eps}}, and the panels in the right column of
figure~{\ref{MWkin.eps}}. The projected mean velocity profile along
the major axis rises more and more slowly as the optical depth
increases, and tends towards a solid body rotation. These trends were
expected, and in agreement with the arguments of Davies~(1990) and the
absorption-only calculations of Bosma et al.~(1992). To quantify this
change of slope, we have calculated, for each optical depth, the major
axis radius at which the mean projected velocity reaches half of its
maximum value $v_\txp^{\text{max}} = 220 \sin i$~km\,s$^{-1}$. These
values are tabulated in the second column of table~{\ref{rot.tab}},
and demonstrate that the effects of dust attenuation are very severe,
even for small optical depths. The reason for this shallower slope of
the rotation curve is that we mainly see the radiation from the
optically thin part of the galaxy, which is more and more restricted
to the outer part of the galaxy as the optical depth increases.

In the same way, we can understand the effects of dust attenuation on
the projected velocity dispersion profile, where also a prominent dust
lane is visible in the contour plot: we mainly see the light (and
hence the Doppler shift) of the stars from the outer part of the disc,
where the intrinsic velocity dispersion is much smaller.

\subsubsection{Galaxies at intermediate inclinations}

Bosma et al.~(1992) found that the signature of dust absorption on the
rotation curve depends strongly on the inclination angle. For
inclinations which deviate as few as 5 degrees from purely edge-on,
they found that the effects of dust on the rotation curve is already
severely reduced, even for optical depths up to $\tau_\txV\sim10$. This
result remains the same if scattering is also taken into account,
because the absorption effects of dust strongly dominate in highly
inclined galaxies. Table~{\ref{rot.tab}} quantitatively demonstrates
these results. For the projected velocity dispersion, a similar effect
is valid, at least for the major axis dispersion profile. When the
entire projected velocity dispersion field (or just the minor axis
dispersion profile) is taken into account, the asymmetrical signature
of the dust can still be recognized for inclinations down to 80
degrees.

Moving to intermediate inclinations, the effects of dust attenuation
on the observed kinematics quickly become very limited. For example,
there is hardly any effect noticeable for an inclination of 60
degrees, which can easily be seen by comparing the corresponding
panels in figure~{\ref{MW_vp.eps}} and~{\ref{MW_sigp.eps}}.

\subsubsection{Face-on galaxies}

In the light of the previous results, we would expect that the effects
of dust attenuation on the observed kinematics of face-on galaxies are
completely neglegible. And indeed, on the mean projected velocity
there is no effect at all: the projected mean velocity is zero for
face-on galaxies, regardless of whether dust is present or
not. However, we do see a conspicuous effect of dust attenuation on
the projected velocity dispersion: it increases with increasing
optical depth. This cannot be an absorption effect, because all stars
along a given line of sight in our face-on galaxy discs have the same
velocity dispersion $\sigma_z$ in the direction of the observer. The
effect arises as a result of photons emitted in the central regions in
a direction parallel to the plane of the galaxy. When these photons
are scattered into a vertical direction, they have a large probability
to leave the galaxy. They will then contribute the high-velocity
information of the stars in the central regions to the LOSVDs, causing
high-velocity wings and hence an increased velocity dispersion.

The effect is similar to what we found for the projected velocity
dispersion profile of elliptical galaxies containing a diffuse dust
component (Baes \& Dejonghe~2002a). Compared to these models, where
the signature of the dust is very significant, it is fairly modest for
the face-on discs. The reason for this weaker impact is the fact that
the velocity dispersion profile in our disc galaxy models have a
fairly shallow slope, compared to the elliptical galaxy models of Baes
\& Dejonghe (2002a). In disc galaxies, the Doppler shifts contributed
by the scattered photons will in the mean differ less from the Doppler
shifts contributed by the direct photons as in elliptical galaxies.

We could ask ourselves the question whether this conclusion is biased
by the chosen star-dust geometry of our disc galaxy models. Indeed,
Baes \& Dejonghe~(2002a) found their results depended strongly on the
relative star-dust geometry: the scattering effect was strong only for
models in which the dust was more extended than the stars. In our disc
galaxy models, the dust distribution has the same scale length as the
stars, whereas the scale height is smaller, such that the dust is
always sandwiched between the stars. Recently however, evidence has
been found for extended distributions of cold dust in spiral galaxies,
both from the radiative transfer modelling of the dust lanes in
edge-on galaxies (Xilouris et al.~1997, 1998, 1999) and from ISO and
SCUBA dust emission (Alton et al.~1999; Davies et al.~1999). In order
to investigate whether this would bias our results, we considered a
new set of models with an extended dust distribution. They were
constructed by increasing the dust scale length or the dust scale
height by factors up to three, while keeping the face-on optical depth
fixed. Running new face-on simulations for these models, we found that
the strength of the scattering effect on the projected velocity
dispersion increases only minimally, with deviations of the order of a
few per cent. The modest effects found in figure~{\ref{MWkin.eps}} are
hence representative for disc galaxies.

\section{Discussion}
\label{discussion.sec}

\subsection{Stellar kinematics in disc galaxies}

One of the most useful and promising applications of stellar
kinematics in disc galaxies is their contribution to mass
modelling. In its simplest form, this mass modelling consists of
decomposing the observed rotation curve into contributions by the disc
and the dark halo (assuming the contributions of a bulge and
molecular/atomic gas are negligible). Even in this most simple
scenario, there is still a virtually complete degeneracy: the
mass-to-light ratio of the disc. Authors have often adopted a solution
in which the contribution of the disc is maximized, leading to the
so-called maximum disc hypothesis (van Albada \& Sancisi 1986). There
are however, no convincing arguments to favour a particular value for
the stellar mass-to-light ratio, and often scenarios in which the dark
matter halo dominates the disc are equally possible (e.g.\ Lake \&
Feinswog 1989). Stellar kinematics can, in principle, provide the
necessary additional constraint to break this degeneracy. Indeed, for
a self-gravitating disc, the vertical velocity dispersion scales
linearly with the square root of the surface density of the disc, and
can therefore be used to constrain the mass-to-light ratio of the
disc.

The most obvious candidates for such an analysis would be face-on
galaxies, because their projected velocity dispersion directly
reflects the vertical velocity dispersion in the disc. We found that
for a dusty disc galaxy, the observed velocity dispersion is affected
by scattering. The effects are modest, however, even for models with
an extended dust distribution. Alas, face-on galaxies do have strong
observational disadvantages. Foremost, their low surface brightness
make it very hard to obtain the stellar kinematics out to sufficiently
large distances to isolate the disc component. Moreover, the shape and
amplitude of the rotation curve cannot be measured directly. The
amplitude can be estimated fairly accurate using the Tully-Fisher
relation, whereas the shape could be approximated through the
(controversial) universal rotation curve formalism (Persic, Salucci \&
Stel~1996). This is unsatisfactory, however, because there is a large
spread in universal rotation curves, and the details of mass modelling
usually depend crucially on small scale features of the rotation curve
which statistical methods cannot incorporate.

Edge-on galaxies don't suffer from these observational problems: their
surface brightness is much higher, and their apparent rotation curve
can be determined directly. However, the interpretation of the
kinematical data is non-trivial. Firstly, the vertical velocity
dispersion must be linked to the observed dispersion, which is a
mixture of contributions from the radial and tangential velocity
components. Based on the observed axis ratio of velocity ellipsoid in
the solar neighborhood, Bottema (1993) estimated the vertical velocity
dispersion in edge-on galaxies by means of the radial velocity
dispersion evaluated at one scale length. It has been shown, however,
that the velocity ellipsoid axis ratios in disc galaxies can vary
substantially (Gerssen et al.~2000). Secondly, our modelling indicates
that both the projected mean velocity and the projected velocity
dispersion field in edge-on galaxies are strongly affected by
interstellar dust.
 
Galaxies of an intermediate inclination ($i<80^\circ$), however, are
most suitable for such an analysis. Gerssen et al.~(1997) have shown
that for galaxies with an intermediate inclination, the three
components of the velocity ellipsoid can, in principle, be determined
by studying the kinematics along major and minor axes. Also, because
the rotation curve of inclined galaxies can directly be determined and
their surface brightnesses are high enough to allow the measurements
of their kinematics in a reasonable time, they form ideal targets for
such a study. Moreover, determining the three components of the
velocity ellipsoid is very useful to constrain the dynamical history
of disc galaxies, because the various proposed mechanisms leave a
different imprint on the velocity ellipsoid. Our modelling
demonstrates that the observed kinematics of inclined disc galaxies
can be directly interpreted and are not biased by dust attenuation. We
argue that studying the observed kinematics of a substantial set of
optically smooth inclined galaxies could thus greatly contribute to
understanding the mass structure and dynamical history of spiral
galaxies.

\subsection{Optical rotation curves in disc galaxies}

The results of our modelling are important concerning the derivation
of the rotation curve of spiral galaxies. Whereas the outer parts of
the rotation curves of spirals were the preferred field of interest
for nearly three decades, it is the inner slope of the rotation curve
which nowadays stands in the spotlight. To measure the slope in this
inner region accurately, a high spatial resolution is required, which
makes the usual 21-cm measurements less suitable, except for the most
nearby galaxies. A useful alternative are CO rotation curves, which
can achieve a high resolution in both the spatial and velocity
directions, but these observations are quite costly in observing
time. The main alternative remains optical H$\alpha$ observations,
either long-slit or Fabry-Perot.

Using H$\alpha$ rotation curves, several authors (Blais-Ouellette,
Amram \& Carignan 2001; Borriello \& Salucci 2001; de Blok, McGaugh \&
Rubin 2001; de Blok \& Bosma 2002) have recently found shallow slopes
for the inner rotation curve of a significant number of (mainly LSB
and dwarf) galaxies. Such slopes are not in agreement with the results
of high-resolution cosmological N-body simulations, where it is found
that dark matter halos have a strong cusp, and hence that rotation
curves of dark matter dominated galaxies should be steeply rising
(Navarro, Frenk \& White 1997; Moore et al. 1997). A number of
``solutions'' have been proposed to explain this discrepancy, amongst
them the suggestion that the internal extinction in the centre of
galaxies could be a major source of uncertainty. Our models show that,
indeed, an edge-on galaxy with an observed slowly rising rotation
curve could in principle be a {\em dusty} galaxy with an intrinsically
steep rotation curve. We found, however, that this effect cannot be
obtained with realistic amounts of dust if the galaxies deviate more
than a few degrees from exactly edge-on, confirming the results of
Bosma et al.~(1992) and Matthews \& Wood (2001). In addition, because
LSB galaxies are assumed to be dust poor (McGaugh~1994, Tully et
al.~1998; Matthews et al.~1999), we have demonstrated that dust
effects cannot explain the observed discrepancies.

We do however demonstrate that dust effects are important for galaxies
within a few degrees from edge-on, and that dust attenuation must be
seriously taken into account in such cases.

\subsection{The importance of the dust grain velocities}


\begin{figure*}
\centering
\includegraphics[clip, bb=43 148 547 493, width=\textwidth]{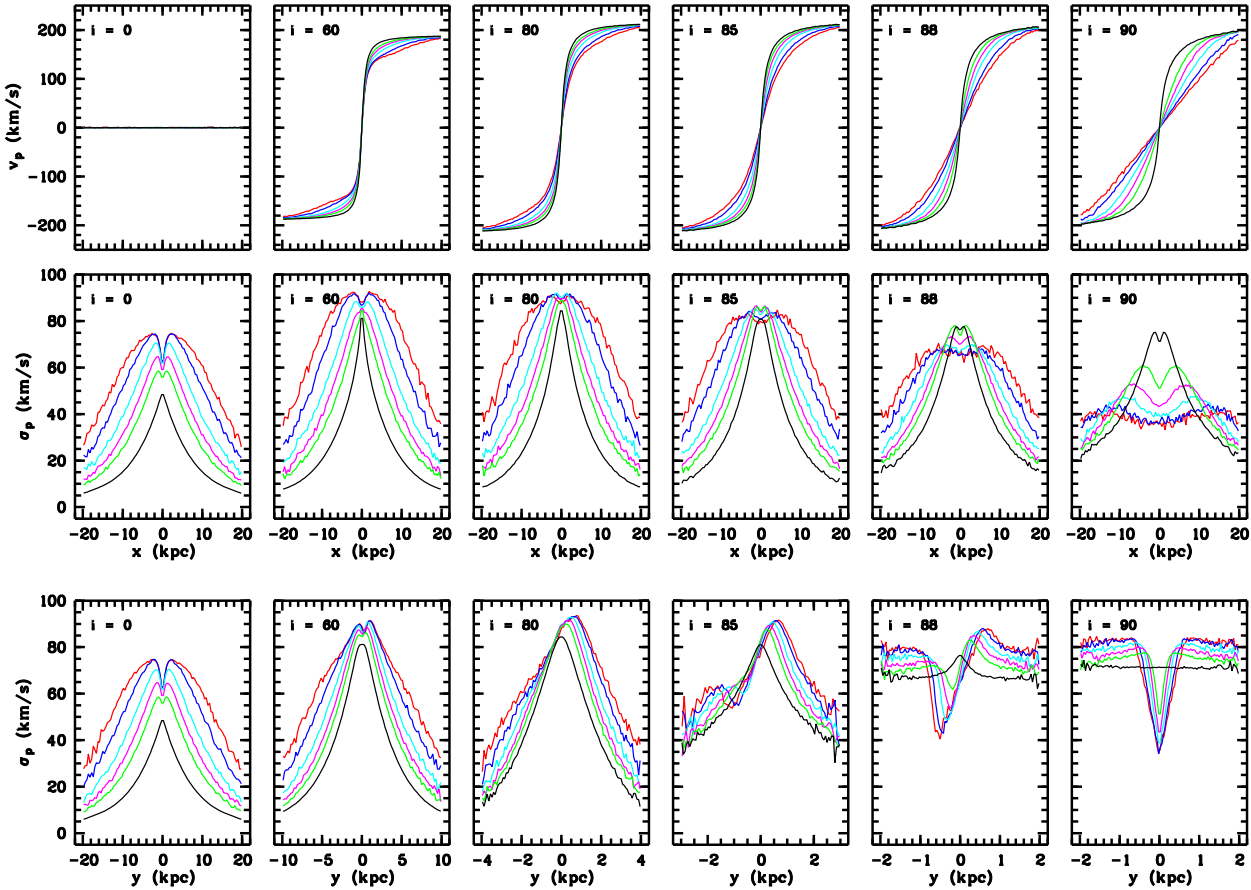}
\caption{The major and minor axes observed kinematics of our disc
galaxy models, calculated without taking the dust velocities into
account. The layout is similar to figure~{\ref{MWkin.eps}}.}
\label{MWappkin.eps}
\end{figure*}

The calculation of the final Doppler shift of a photon in the SKIRT
code takes into account the velocity information of the star and the
individual velocity vectors of each individual scattering dust
grain. This means that for every scattering event, we need to sample a
velocity from the dust velocity field, and this is a costly
operator. It would be very convenient from a computational point of
view, if we could neglect the dust grain velocities. We have done this
in our previous Monte Carlo simulations (Baes \& Dejonghe 2001b,
2002a), where we studied the observed stellar kinematics of elliptical
galaxies. In these models, both stars and dust grains are supported by
random motions, and because the dust is colder than the stars, the
extra Doppler shifts due to scattering are generally smaller than the
stellar Doppler shift. In a spiral galaxy, however, this argument does
not hold anymore, because the velocities of both stars and dust grains
are dominated by the rotation and are therefore of the same magnitude.

In order to investigate the importance of including the individual
dust grain velocities into the code, we ran the same models again but
without taking the dust velocities into account. In
figure~{\ref{MWappkin.eps}} we plot the major and minor axes mean
projected velocity and projected velocity dispersion profiles of our
models calculated this way. Comparing this figure with the
corresponding this figure with figure~{\ref{MWkin.eps}}, we find that
the effects of dust on the mean projected velocity profiles are
significantly overestimated when the dust grain velocities are not
taken into account, in particular for inclined galaxies. The
differences between the projected velocity dispersion curves in both
figures are even worse: when the dust grain velocities are neglected,
the apparent effect of dust attenuation is to increase the dispersion
at nearly all radii and all inclinations, by factors of up to 200 per
cent or more for realistic optical depths.

To understand this behaviour, consider a photon emitted by a star with
velocity $\bfv_*$ in the direction $\bfk_0$, and assume this photon is
scattered by a dust grain with velocity $\bfv_\txd$ into the direction
$\bfk_1$, whereafter it manages to escape the galaxy. When we do not
take the dust grain velocity into account, the observed Doppler shift
of the photon will simply be $u_0=\bfv_*\cdot\bfk_0$. On the contrary,
the correct observed Doppler shift of the photon reads
\begin{align}
	u_1
	&=
	\bfv_*\cdot\bfk_0
	+
	\bfv_\txd\cdot(\bfk_1-\bfk_0) \\
	&=
	(\bfv_*-\bfv_\txd)\cdot\bfk_0 + \bfv_\txd\cdot\bfk_1.
\end{align}
In a disc galaxy, the motion of both stars and dust grains is
dominated by the same rotation. If the pathlength of the photon
between its emission and scattering is short, the velocity vectors of
the star and dust grain will therefore be very similar,
$\bfv_*\approx\bfv_\txd$, such that $u_1 \approx
\bfv_*\cdot\bfk_1$. The new correct Doppler shift is hence a typical
line-of-sight velocity of a star in the new propagation direction
$\bfk_1$ of the photon. Therefore, taking the dust motion into account
more or less forces the dust grain to adopt a Doppler shift that is
``appropriate'' for the new propagation direction. Along a given
line-of-sight, the velocity information carried by scattered photons
will, in the mean, only modestly deviate from the velocity information
carried by photons directly emitted in same direction. If we don't
take the dust grain velocities into account, the photon can
contribute, to the observed kinematics in a given direction, a
line-of-sight velocity typical for a completely different direction.

In the elliptical galaxy models from Baes \& Dejonghe (2002a), this
does not make much difference, because the velocities of the stars
have similar magnitudes in the different directions. In a rotating
disc galaxy however, this makes a huge difference because the stars
have hugely different velocities in different directions: the vertical
and radial motions are determined by the dispersion and are of the
order of a few tens of km/s, whereas the azimuthal motion is dominated
by the rotation and is of the order of 200 km/s. If the dust grain
velocities are not correctly taken into account, scattered photons can
thus cause erroneous extremely-high-velocity wings in the
LOSVDs. These give rise to modest but significant errors in the mean
projected velocity, and very large errors in the projected velocity
dispersion.

As a remark, we want to point out that Matthews \& Wood (2001) found
that ``also including the Doppler shifts arising from the relative
bulk motion of the scattering dust particles had a negligible effect
on the final rotation curves''. Whereas this statement is indeed true
for the mean projected velocity of highly inclined galaxies, we
nevertheless judge that this could easily be misinterpreted as a
justification to neglect dust velocities altogether. It must be
realized that for the calculation of the entire LOSVDs, and for the
projected velocity dispersion in particular, the dust velocities do
play a crucial role, and must be taken into account.

\section{Conclusions}
\label{conclusions.sec}

In this paper we have presented a novel Monte Carlo code that can take
velocity information into account, and can hence be used for
kinematical modelling of dusty objects. We applied this code to
calculate the observed kinematics of dusty disc galaxies. The main
results of this paper can be summarized as follows.
\begin{enumerate}
\item A correct inclusion of kinematical information into radiative
transfer problems requires the inclusion of the velocities of both the
emitting stars and the scattering dust grains. We see no other way of
tackling this problem except with Monte Carlo techniques.
\item A new approach is presented to optimize the integration through
the dust in a Monte Carlo code, using a trilinear interpolation
instead of a constant opacity within each cell. We compared both kinds
of grids, and find that, to obtain a similar accuracy, the new
approach is more efficient in terms of computation time and memory.
\item The effects of dust attenuation on the kinematics of edge-on
disc galaxies are severe. Both the mean projected velocity and the
projected velocity dispersion are severely affected, even for modest
optical depths. Therefore we strongly advise to always interprete the
stellar kinematics and optical rotation curves of edge-on galaxies
very cautiously.
\item For galaxies which are more than a few degrees from strictly
edge-on, the effects of interstellar dust on the observed kinematics
are much weaker. Therefore, we argue that dust attenuation cannot be
invoked as a possible mechanism to reconcile the observed slope of LSB
galaxies with the predicted CDM cosmological models.
\item Dust attenuation does not affect the kinematics of
intermediately inclined galaxies. Such galaxies hence form the ideal
targets for stellar kinematical studies, in particular to constrain
the mass structure and to study the kinematical history of disc
galaxies.
\item The projected velocity dispersion of face-on galaxies increases
slightly due to scattering of dust grains into the line of sight. The
effects are relatively small, however, even for extended dust
distributions.
\item Neglecting the extra Doppler effect caused by the scattering
medium results in incorrect projected kinematics, in particular if the
velocity components of the stars (and dust grains) in the galaxy
differ considerably, such as in a rotating disc. If the dust grain
velocities are neglected, the mean projected velocity is significantly
underestimated, and the projected velocity dispersion field is
completely overestimated.
\end{enumerate}

\section*{Acknowledgements}

M.\,B.\ gratefully thanks the Fund for Scientific Research for
financial support.

\appendix
\section{Generating positions from an exponential disc galaxy}

In this Appendix, we show how the SKIRT code generates a random
position from the exponential disc model, i.e.\ how a random position
is drawn from the three-dimensional probability density
\begin{equation}
	p(\bfr) \txd\bfr
	\propto
	\exp\left(-\frac{R}{h_*}\right)
	\exp\left(-\frac{|z|}{z_*}\right)
	\txd\bfr.
\end{equation}
As this probability density is a separable function of $R$, $\varphi$
and $z$, we can generate a random position by independently generating
each of these three coordinates, whereby the generation of an azimuth
is trivial. To generate a random radius and height, we use the
transformation technique, which says that a random variable $x$ can be
drawn from a probability density $p(x)\txd x$ by generating a uniform
deviate $X$ and solving the equation
\begin{equation}
	X = \int_{-\infty}^x p(x')\,\txd x'
	\left/
	\int_{-\infty}^\infty p(x')\,\txd x'
	\right.
\end{equation}
for $x$. The calculation of a random $z$ is easy using this principle
and results in
\begin{equation}
	z
	=
	z_*
	\sgn(1+2X)\ln(1-|1-2X|).
\end{equation}
This procedure works equally well when we introduce a cut-off, or when
a sech or isothermal profile is used instead of the exponential
profile to describe the vertical distribution of the stars
(e.g.~Bianchi et al.~1996).

Applying the same technique to generate a random $R$, we can find $R$
by solving the equation
\begin{equation}
	X 
	= 
	1-\left(1+\frac{R}{h_*}\right)
	\exp\left(-\frac{R}{h_*}\right).
\end{equation}
Bianchi et al.\ (1996) invert this transcendental equation numerically
to find $R$. However, this equation can be solved exactly by means of
the Lambert function, yielding
\begin{equation}
      	R 
	= 
	h_* \left[ 
	-1-W_{-1}\left(\frac{X-1}{\text{e}}\right) 
	\right].
\end{equation} 
The Lambert function, also known as the product log function, is
generally defined as the inverse of the function $w \rightarrow f(w) =
w\,{\text{e}}^w$, and $W_{-1}(z)$ represents the only real branch of this
complex function besides the principle branch. The Lambert function
can be computed in a very efficient way by means of Halley iteration;
for more information on this function see Corless et al.~(1996).

\bsp

\end{document}